\documentclass[twocolumn,showpacs,preprintnumbers,amsmath,amssymb,showkeys]{revtex4}
\usepackage{bm}
\usepackage[latin1]{inputenc}

\newcommand{\ii}{\'{\i}}
\newcommand{\tr}{{\rm Tr}\,}
\begin{document}

\title{
Geometrical aspects of a generalized statistical mechanics
}

\author{M.~Portesi$^1$}~\email{portesi@fisica.unlp.edu.ar}
\author{F.~Pennini$^{1,2}$}
\author{A.~Plastino$^1$}

\affiliation{
$^1$~Instituto de F\'{\i}sica La Plata (IFLP, CONICET--UNLP) \ and \\
\mbox{Dpto.~F\ii sica, Fac.~Cs.~Exactas, Univ.~Nacional de La Plata} \\
C.C.~67, 1900 La Plata, Argentina \\
$^2$~Departamento de F\'{\i}sica, Univ.~Cat\'olica del Norte, \\
Av.~Angamos 0610, Antofagasta, Chile
}

\date{\today}

\begin{abstract}

We discuss here the use of generalized forms of entropy, taken as
information measures, to characterize phase transitions and critical
behavior in thermodynamic systems. Our study is based on geometric
considerations pertaining to the space of parameters that describe
statistical mechanics models. The thermodynamic stability of the system is
the focus of attention in this geometric context.

\end{abstract}



\maketitle

\section{Introduction}
\label{section_introduction}

Our aim is to add a note on current efforts to investigate the geometrical
structure of the space of thermodynamic parameters for classical and quantum
systems. This kind of studies has been developed by many authors (see, for
instance, Refs.~\cite{%
w_jcp63,r_pra20rmp67,m_p125a,i_tns30,j_rmp24,jm_pra39,j_jpa23,br_pre51,jjk}),
and has been applied to the ideal gas, the van der Waals gas, and magnetic
systems that exhibit phase transitions. In such instances the Riemannian
scalar curvature $R$ of the parameters' space plays such an important role
in detecting critical behavior that it has been interpreted as a measure of
the stability of the thermodynamic system (typically, for non-interacting
models one obtains $R=0$, i.e.\ a flat geometry, while $R$ diverges at the
critical point for interacting systems).

In Ref.~\onlinecite{w_jcp63} Weinhold considered the thermodynamic surface
given by the fundamental relation $U=U(\{X_i\})$ in the $(r+1)$-dimensional
Gibbs space (with coordinates labelled by $U,X_1,\ldots,X_r$), and obtained
the components of the metric tensor of that space as the second derivatives
of the internal energy $U$ with respect to each pair of the $r$ extensive
parameters $X_i$. Ruppeiner \cite{r_pra20rmp67} focused attention on
fluctuations of the thermodynamic magnitudes and obtained the metric tensor
via second moments of the fluctuations. The two ensuing metrics have been
proven by Mruga{\l}a to be equivalent \cite{m_p125a}. Another statistical
path for reaching a Riemannian metric \cite{lecorbeiller,santalo,poole} in
the space of thermodynamic parameters is that originated in the works by Rao
\cite{r_bcms37} and Amari (see, for instance, \cite{amarian_2000}) in the
field of statistical mathematics, and also by Ingarden \cite{i_tns30},
Janyszek \cite{j_rmp24}, and other authors in the field of thermodynamics
and statistical mechanics. The concomitant information-theoretic approach is
based on the concept of relative entropy and given in terms of the
Boltzmann--Gibbs--Shannon entropy.

There are also some recent efforts in the field of information geometry
related with the generalized, nonextensive formulation of statistical
mechanics \cite{t_jsp52,tsallisURL}. Among these studies one finds the
contributions by Amari, Nagaoka and coworkers (\cite{amarian_2000} among
others) in connection with the geometrical structure in the manifold of
probability distributions, and the analysis by Abe \cite{a_pre68} of the
geometry of escort distributions, in connection with bit-variance and
fluctuations of the crowding index of a multifractal. A new geometrical
approach to thermo-statistical mechanics is introduced in~\cite{tb_0203536},
where the relevance of the approach within the contexts of nonextensive
statistical thermodynamics is analyzed, showing that Riemannian geometry
concepts yield a powerful tool. More recently, Naudts \cite{n_osid12}
studies escort density operators and generalized Fisher information
measures.

We will analyze here the geometrical approach to statistical
mechanics with regard to generalized measures of information, as
the one usually denoted as Tsallis' nonextensive
$q$-entropy~\cite{t_jsp52,tsallisURL}
\begin{equation}
S_q(\hat\rho)\equiv k_B\frac{1-{\rm Tr}\,\hat\rho^q}{q-1} \qquad (q\in\mathbb{R})
\label{Sq} \end{equation}
for which we will show how to derive the metric tensor and other
geometric quantities in the space of parameters that define the
state of the system, $\hat\rho$. Our considerations will refer to
a generalized thermo-statistical context characterized by a given
nonextensive index $q$ and by a given prescription for the
expectation values of relevant magnitudes.

The paper is organized as follows: in Section~\ref{section_olm} we
summarize some basic results for the density operator in a
generalized statistical framework. Section~\ref{section_infogeom1}
is devoted to exhibit some standard results on the geometrical
aspects of the space of parameters, while in
Section~\ref{section_infogeomq}we present an extension of that
study to a generalized statistical framework. For an arbitrary
value of the nonextensivity index $q$, we obtain the metric tensor
of the parameters' space and discuss the meaning of various
alternative ways of rewriting the metric tensor. Then, we
calculate the scalar curvature in terms of the pseudo-partition
function (given by Eq.~(\ref{Zqraya}) below), the generalized mean
values, and their derivatives with respect to the thermodynamic
parameters. As an application we consider the ideal gas case in
Section~\ref{section_idealgas}. Finally some conclusions are
remarked in Section~\ref{section_conclusions}.

\section{The density operator for generalized statistics' OLM-version}
\label{section_olm}

The density operator that maximizes Tsallis nonextensive
$q$-entropy~(\ref{Sq}) for a given positive value of $q$ is written, within
the optimized Lagrange multipliers (OLM) formalism, as~\cite{mnpp_pa286}
\begin{equation}
\hat\rho=\frac{1}{\bar Z_q} \; e_q\!\left(-\sum_{i=1}^r \beta^i
(\hat F_i-m_i)\right)
\label{rhoq}
\end{equation}
where $m_1,\ldots,m_r$ are the generalized expectation values of
$r$ quantum operators. They are considered to be known as prior
information and are given by the following generalized
prescription~\cite{mnpp_pa286,tmp_pa261}
\begin{equation}
m_i=\langle\hat F_i\rangle_q\equiv
\frac{{\rm Tr}\,\hat\rho^q \hat F_i}{{\rm Tr}\,\hat\rho^q} \,
, \qquad i=1,\ldots,r
\label{qmeanvalue}
\end{equation}
(for the sake of notational simplicity, we do not use along this
article any label $q$ for $\hat\rho$ nor $m_i$ despite the fact
that they do depend on that index; we keep, however, the subindex
in quantities such as $S_q$ and $\bar Z_q$).
The set of Lagrange multipliers $\{\beta^1,\ldots,\beta^r\}$ is
such that it fits the restrictions~(\ref{qmeanvalue}) in the
procedure of constrained extremization of $S_q$ when one is
working {\it within the OLM formalism}~\cite{mnpp_pa286}, i.e.\
the restrictions are rephrased as centered mean quantities, ${\rm
Tr}[\hat\rho^q (\hat F_i-m_i)]=0$. This formalism is an
alternative to the TMP one of Ref.~\onlinecite{tmp_pa261}. It has
been established~\cite{ampp_pla281,a_pa300} that, within OLM
method, the $\{\beta^i\}$ correspond to the {\it physical}
intensive parameters (e.g.\ the multiplier associated with the
Hamiltonian operator, $\hat F_1=\hat H$, is
$\beta^1=\beta=1/(k_BT)$ where $T$ is the physical temperature of
the system). In this framework, the pseudo-partition function
\begin{equation}
\bar Z_q\equiv{\rm Tr}\,e_q\!\left(-\sum_{i=1}^r \beta^i (\hat
F_i-m_i)\right)
\label{Zqraya}
\end{equation}
is defined in such a form that the density operator is automatically
normalized, i.e.\ ${\rm Tr}\,\hat\rho=1$. Notice that no Lagrange
multiplier associated with this normalization condition for $\hat\rho$ is
explicitly mentioned in writing down the equilibrium density
matrix; instead, here we choose to introduce the pseudo-partition function in
the expression for $\hat\rho$. In all these expressions, $e_q(x)$
stands for the $q$-exponential function
\begin{equation}
e_q(x)\equiv[1+(1-q)x]^{1/(1-q)}_+
\end{equation}
with $[X]_+\equiv X\,\Theta(X)$, being $\Theta$ the Heaviside step function.
Finally, the ``extensive limit"
corresponds to the situation $q\rightarrow 1$, and $q-1$
becomes a measure of the degree of entropic nonextensivity.

\section{Geometrical aspects of parameters' space in standard statistics}
\label{section_infogeom1}

One of the key ideas for achieving our aims is to represent the
equilibrium density operator, and related quantities, in an
$r$-dimensional space whose coordinate axes correspond to the
parameters $\beta^1,\ldots,\beta^r$ (called the statistical
temperatures). These are associated to the operators $\hat
F_1,\ldots,\hat F_r$, the mean values of which are assumed to
constitute the known prior information.
One can consider ``neighbor" quantum states in the $r$-dimensional space of
parameters and, from a notion of distance between states, derive the metric
tensor. Janyszek and other
authors~\cite{jm_pra39,j_jpa23,r_pra20rmp67,br_pre51,jjk} have done so
for the equilibrium density operator that maximizes von
Neumman--Shannon--Gibbs entropy, i.e.\
$\hat\rho=Z^{-1}\,\exp(-\sum\beta^i\hat F_i)$ \ with
$Z=\tr\exp(-\sum\beta^i\hat F_i)$. For the case of commuting operators $\hat
F_i$, the metric tensor can be defined as
\begin{equation}
g_{ij}(\beta)=
\langle\partial_j\ln\hat\rho\ \, \partial_i\ln\hat\rho\rangle \ ,
\qquad i,j=1,\ldots,r
\label{gij1sim}
\end{equation}
where $\langle\ldots\rangle\equiv\tr(\hat\rho\,\ldots)$ \ and \
$\partial_i\equiv\frac{\partial\ }{\partial\beta^i}$. For the case
of non-commuting operators, this expression should at least be
symmetrized. However a different derivation can be adopted,
introducing the notion of information distance between two quantum
states, say $\hat\rho$ and $\hat\sigma$, through the relative
entropy or information gain (also known as Kullback--Leibler
divergence), which is given by
\begin{equation}
K(\hat\rho\|\hat\sigma)=\tr\hat\rho(\ln\hat\rho-\ln\hat\sigma)
\label{K}
\end{equation}

The derivation of the metric tensor is as follows (see, for
instance, \cite{jm_pra39}): first a symmetrical (squared)
information distance is constructed, as the sum
$D^2(\hat\rho,\hat\sigma)\equiv
K(\hat\rho\|\hat\sigma)+K(\hat\sigma\|\hat\rho)$; this quantity is
evaluated for a pair of equilibrium density operators,
$\hat\rho(\{\beta\})$ and $\hat\rho(\{\beta+\Delta\beta\})$, which
are neighbor states in the parameters' space. The ensuing
expression is then expanded as a Taylor series for small
displacements $\Delta\beta^i$ in all directions, resulting in
vanishing zeroth and first order contributions. The non-vanishing
second order terms thus allow for the definition of the metric
tensor from the local square distance $dD^2=g_{ij}\,d\beta^i
d\beta^j$ as
\begin{eqnarray}
g_{ij}(\beta) & = &
-\langle\partial_j\partial_i\ln\hat\rho\rangle
\label{gij1taylorRho} \\
& = &
\partial_j\partial_i\ln Z
\label{gij1taylorZ}
\end{eqnarray}
or, recalling that $\langle\hat F_i\rangle=-\partial_i\ln Z$,
\begin{equation}
g_{ij}(\beta)=-\partial_j \langle\hat F_i\rangle
\label{gij1mj}
\end{equation}

For classical systems or commuting operators, the definitions for
$g_{ij}$ given by Eqs.~(\ref{gij1sim}) and~(\ref{gij1taylorRho})
do coincide. In these cases it is also seen that the fundamental
tensor is related to thermodynamic fluctuations, as the component
$ij$ of the tensor corresponds to the covariance or second moment
of the pair of operators $\hat F_i$ and $\hat F_j$:
\begin{equation}
g_{ij}=
\langle\,(\hat F_i-\langle\hat F_i\rangle)
(\hat F_j-\langle\hat F_j\rangle)\,\rangle
\label{gij1fluc}
\end{equation}
Let us comment that for non-commuting operators an integral expression for the
covariances is
introduced, based on this connection with the tensor $g_{ij}$
--see Ref.~\onlinecite{jm_pra39} for
details--. Finally, we notice that the metric tensor, being
constructed from second derivatives of the Kullback--Leibler
divergence, corresponds to the Fisher information matrix.

The metric tensor is one of the key ingredients in the geometric
approach to thermostatistics. All the different viewpoints
summarized here, Eqs.~(\ref{gij1taylorRho})--(\ref{gij1fluc}),
have been worked out by many authors in the framework of standard
($q=1$) statistics, and a number of applications to different
physical models have been analyzed in this way. We present in the
following section the formalism leading to analogous results for
generalized statistics.

\section{Extension of geometrical concepts to nonextensive statistical
frameworks}
\label{section_infogeomq}

\subsection{The fundamental tensor}

In the wake of the approach presented in Ref.~\onlinecite{jm_pra39} and
working here along similar lines, we consider a generalization of the
definition of information gain to nonextensive settings of index $q\neq 1$.
The quantum $q$-divergence~\cite{lmpr_pa334,a_pra68,a_pa344} for two
normalized operators $\hat\rho$ and $\hat\sigma$ has been defined as
\begin{subequations}
\label{Kq}
\begin{equation}
K_q(\hat\rho\|\hat\sigma) =
\frac{1}{1-q} \, \left(
\tr\hat\rho
-\tr\hat\rho^q \hat\sigma^{1-q}\right)
\label{Kq-1}
\end{equation}
generalizing the quantum Kullback--Leibler divergence or relative
entropy, Eq.~(\ref{K}). $K_q$ can also be written in the
following equivalent form,
\begin{equation}
K_q(\hat\rho\|\hat\sigma)=\tr\hat\rho^q(\ln_q\hat\rho-\ln_q\hat\sigma)
\label{Kq-2}
\end{equation}
\end{subequations}
where the so-called $q$-logarithm \ $\ln_q x \equiv \frac{x^{1-q}-1}{1-q}$ \
for $x\geq 0$ (excluding $x=0$ if $q>1$), represents the inverse function of
$e_q(x)$. It is straightforward to show that $e_q(x)$ and $\ln_q(x)$ tend to
the exponential and logarithm functions, respectively, for $q$ sufficiently
near 1 (standard limit).

The next step in our derivation is to extend the concept of (squared)
information distance, $D^2(\hat\rho,\hat\sigma)$. One can see that
$K_q(\hat\rho\|\hat\sigma)\neq K_q(\hat\sigma\|\hat\rho)$, except for the
particular cases: (i) $q=1/2$ (Hellinger distance, with the conditions
$\tr\hat\rho=\tr\hat\sigma=1$), and (ii) $q=0$ (since $K_0=0$ for any pair
of normalized density operators). In general, then, for our purposes we make
use of the symmetrized sum
\begin{equation}
K_q(\hat\rho\|\hat\sigma)+K_q(\hat\sigma\|\hat\rho)
\label{Dq2}
\end{equation}
as information measure. From
a Taylor expansion of this quantity around a certain set of values for the
parameters $\{\beta_i\}$ (i.e., a given point in the $r$-dimensional
$\beta$-space), that characterize the state of the system,
we determine the
form of the fundamental tensor in a generalized context
(in fact, we determine a {\it family} of
metrics which are characterized by the single index $q$, with $q\in\mathbb{R}^+$). Let us
then construct the first few terms of the associated Taylor
series, taking $\hat\rho=\hat\rho(\beta)$ and $\hat\sigma=\hat\rho(\beta+\Delta\beta)$:
\begin{itemize}
\item
The zeroth-order term is identically zero since
$K_q(\hat\rho\|\hat\sigma)=0$ if and only if
$\hat\rho=\hat\sigma$. This was also the case for the standard
($q=1$) context.
\item
{}From Eqs.~(\ref{Kq}) and~(\ref{Dq2}), the first-order term vanishes
identically as in the standard case. This is a consequence of the shape
adopted for the symmetrized information gain. In other words, the explicit
form of the density operator does not play any role at this level.
\item From the second-order terms in our Taylor expansion of the
symmetric information measure~(\ref{Dq2}), we obtain the following
result, which we interpret as the {\it generalized fundamental
tensor within the nonextensive picture}:
\begin{eqnarray}
g^{(q)}_{ij} & = &
q\,\tr\left(\hat\rho^{-1}\frac{\partial\hat\rho}{\partial\beta^j}
\frac{\partial\hat\rho}{\partial\beta^i}\right)
\label{gqij11} \\
& = & -q\,\tr\left(\hat\rho\,\frac{\partial^2\ln\hat\rho}
{\partial\beta^j\,\partial\beta^i}\right)
\label{gqij2}
\end{eqnarray}
where $\hat\rho(\beta)$ is the OLM density matrix~(\ref{rhoq}).
We notice that this result can also be regarded as the $ij$ element of the
{\it generalized Fisher information~(GFI) matrix} \
$G^{(q)}=\left(g_{ij}^{(q)}\right)$ \ (see, for instance,~\cite{tth_aj480}
and references therein). Interestingly enough, from Eqs.~(\ref{gij1sim}),
(\ref{gij1taylorRho}), (\ref{gqij11}) and (\ref{gqij2}) we gather that this
quantity \mbox{--when} expressed as a trace-- exhibits shape invariance,
irrespective of the value of $q$. Indeed, the form of the GFI matrix is seen
to be the same for any value of the nonextensivity index~$q$, and it also
coincides with that of the standard case. Differences arise, of course, when
one introduces explicitly the actual density operator, which may correspond
either to the extensive or to the nonextensive scenario.
\end{itemize}

Summing up, we have found the fundamental tensor of the space of
parameters within a nonextensive statistical framework, in terms
of first logarithmic derivatives of the generalized density
operator, $g_{ij}^{(q)}=q\tr(\hat\rho\;\partial_j\ln\hat\rho\;
\partial_i\ln\hat\rho)$, as well as in terms of second derivatives
of $\ln\hat\rho$, Eq.~(\ref{gqij2}), with $\hat\rho$ given as in
Eq.~(\ref{rhoq}). Notice that, in both expressions, the {\it standard}
logarithm function is involved.

It is instructive to look at alternative formulations of the
results already obtained. After a little algebra we can rewrite
each of the expressions mentioned previously, in terms of
variations of the logarithm of the pseudo-partition function:
\begin{eqnarray}
g_{ij}^{(q)} & = & q\{-\partial_j m_i- (q-1)\bar
Z_q^{2(q-1)}\tr[\hat\rho^{2q-1} \times
\nonumber\\
&& \quad \times (\delta_q\hat F_j-\partial_j\ln\bar Z_q)
(\delta_q\hat F_i-\partial_i\ln\bar Z_q)]\}
\end{eqnarray}
and
\begin{eqnarray}
g_{ij}^{(q)} & = & q\{-\partial_j\ln\bar Z_q\,\partial_i\ln\bar Z_q +
\bar Z_q^{2(q-1)}\tr[\hat\rho^{2q-1} \times
\nonumber \\
&& \quad \times (\delta_q\hat F_j-\partial_j\ln\bar Z_q)
(\delta_q\hat F_i-\partial_i\ln\bar Z_q)]\}
\end{eqnarray}
respectively, where the generalized deviation operators are
defined as $\delta_q\hat F_i\equiv\hat F_i-m_i$ with $m_i$ given
by Eq.~(\ref{qmeanvalue}). Comparing these expressions we
immediately obtain a useful (symmetric) relation for the first
derivative of a given mean value with respect to any of the
Lagrange parameters that writes
\begin{eqnarray}
\partial_j m_i=
\partial_i m_j & = &
\partial_j\ln\bar Z_q\,\partial_i\ln\bar Z_q -q
\bar Z_q^{2(q-1)}\tr[\hat\rho^{2q-1} \times
\nonumber \\
&& \!\!\times (\delta_q\hat F_j-\partial_j\ln\bar Z_q)
(\delta_q\hat F_i-\partial_i\ln\bar Z_q)]
\label{djmi}
\end{eqnarray}
from which the components of the metric tensor can finally be cast in compact form as
\begin{equation}
g^{(q)}_{ij}=-\partial_j m_i-(q-1)\partial_j\ln\bar Z_q\,\partial_i\ln\bar Z_q
\label{gqijZ}
\end{equation}
This covariant tensor of rank 2 is seen to be symmetric (by use of
Eq.~(\ref{djmi})~). Its conjugate or reciprocal tensor,
$g^{(q)\,ij}$, is a symmetric tensor of contravariant valence two.

Yet another form of representing the $q$-metric tensor involves
the use of fluctuations, as in Eq.~(\ref{gij1fluc}). This point is
discussed with some detail in Ref.~\onlinecite{0510434}, where
generalized expressions for operator variance and covariance are
also considered.

We see from (\ref{gqijZ}) that, in order to evaluate the
generalized metric tensor for a given system, we need the
logarithmic derivatives of the pseudo-partition function with
respect to the Lagrange parameters, as well as the behavior of the
generalized expectation values of the relevant operators in
$\beta$-space. We end up by remarking that the standard limit for
the fundamental tensor is correctly obtained in all the equivalent
forms developed in this section.

\subsection{The scalar curvature}

Let us now address the computation of the curvature of the
Riemannian space of parameters. We remind that in the extensive
$q=1$ framework~\cite{jm_pra39} the metric tensor $G=(g_{ij})$
corresponds to the matrix of second partial derivatives of the
function $f(\{\beta\})=\ln Z$ with respect to every pair of
parameters, i.e.\ the Hessian matrix of $\ln Z$, $G=H(\ln Z)$ (see
Eq.~(\ref{gij1taylorZ})~). As a consequence, the curvature tensor
and the scalar curvature of the Riemann space, being constructed
from derivatives of the elements of $G$, adopt a relatively simple
form. Using Eq.~(\ref{gij1taylorZ}), the results are the
following~\cite{jm_pra39}:
\begin{equation}
R_{ijkl}=\frac 14\, g^{mn}(f_{,mil}f_{,njk}-f_{,mik}f_{,njl}) \, ,
\qquad f\equiv \ln Z
\label{R_ijkl}
\end{equation}
for the curvature tensor, where $f_{,i}\equiv\partial_i
f=\partial\ln Z/\partial\beta^i$ and so on, and
\begin{equation}
R=\frac 2g\, R_{1212} \, , \qquad g=|G|
\label{R}
\end{equation}
for the scalar curvature, with $g=|H(f)|$. Equation~(\ref{R}) is valid in
the particular case of considering only $r=2$ parameters.

In a nonextensive setting characterized by a given fixed value of
$q$, we arrive at more involved expressions that, however, reduce
to the corresponding $q\rightarrow 1$ standard limit. For the sake
of brevity, we define here
\[
\bar f\equiv\ln\bar Z_q
\]
(using the {\it standard} logarithm) so that the
Riemann--Christoffel tensor reads~\cite{santalo,lecorbeiller}
\begin{widetext}
\begin{eqnarray}
R^{(q)}_{ijkl}&=&
-(q-1)(\bar f_{,ik}\bar f_{,jl}-\bar f_{,il}\bar f_{,jk})
-g^{(q)\,st}\,\bigg\{
\frac 14(m_{s,jl}m_{t,ik}-m_{s,jk}m_{t,il}) + \nonumber \\
&&+\frac 12 (q-1) \left[ \bar f_{,s} (m_{t,ik}\bar f_{,jl}-m_{t,il}\bar f_{,jk})
+(\bar f_{,ik}m_{s,jl}-\bar f_{,il}m_{s,jk}) \bar f_{,t}\right]
+(q-1)^2 (\bar f_{,ik}\bar f_{,jl}-\bar f_{,il}\bar f_{,jk}) \bar f_{,s} \bar f_{,t}
\bigg\}
\label{Rq_ijkl}
\end{eqnarray}
\end{widetext}
(with the notation $m_{s,ij}\equiv\partial_j\partial_i m_s$). This is one of
our main results in the present effort. The $m_i$ correspond to {\it
generalized} expectation values, and $\bar f$ stands for the logarithm of
the {\it pseudo}-partition function. Apart from those $q$-dependent
quantities, we appreciate in Eq.~(\ref{Rq_ijkl}) the fact that \ (i)~some
new terms, of order $(q-1)$ and $(q-1)^2$, emerge in the nonextensive
framework with respect to the standard result (\ref{R_ijkl}), and \ (ii)~the
metric tensor itself contains an explicit contribution with $(q-1)$, arising
from Eq.~(\ref{gqijZ}). It can be checked that the only terms that survive
on letting $q\rightarrow 1$, exactly correspond to Eq.~(\ref{R_ijkl}).

We can now compute, by appropriate contractions, first the Ricci tensor
$R^{(q)}_{jk}=g^{(q)\,il}\,R^{(q)}_{ijkl}$ and then the scalar curvature
$R^{(q)}=g^{(q)\,jk}\,R^{(q)}_{jk}$. A combinatorial analysis shows that,
for an $r$-dimensional Riemann space, the number of independent components
of the $(4,0)$-type Riemann--Christoffel curvature tensor is
$r^2(r^2-1)/12$, due to symmetry reasons~\cite{santalo}. At this point, in
order to fix ideas and also for application purposes, we address the
particular case of $r=2$ Lagrange parameters characterizing the system. In
this case it suffices merely to obtain an expression for $R^{(q)}_{1212}$
and, then, the desired scalar curvature $R^{(q)}$ can be computed making use
of a relation of the form given in Eq.~(\ref{R}) (notice that this equation
is valid for {\it any} 2-dimensional metric). On the one hand, we calculate
the $2\times 2$-determinant $g^{(q)}\equiv|G^{(q)}|$:
\begin{eqnarray}
g^{(q)} &  = & \left|\left(\begin{array}{cc}
m_{1,1} & m_{1,2} \\ m_{2,1} & m_{2,2}
\end{array}\right)+(q-1)\left(\begin{array}{cc}
\bar f_{,1}^2 & \bar f_{,1} \bar f_{,2} \\ \bar f_{,2} \bar f_{,1} & \bar f_{,2}^2
\end{array}\right)\right| = \nonumber \\
& = &\left|\begin{array}{cc}
m_{1,1} & m_{1,2} \\ m_{2,1} & m_{2,2}
\end{array}\right| + (q-1) \times \nonumber \\
&& \ \times \left\{\bar f_{,1} \left|\begin{array}{cc}
\bar f_{,1} & \bar f_{,2} \\ m_{2,1} & m_{2,2}
\end{array}\right|-\bar f_{,2} \left|\begin{array}{cc}
\bar f_{,1} & \bar f_{,2} \\ m_{1,1} & m_{1,2}
\end{array}\right|\,\right\}
\label{detgq}
\end{eqnarray}
and, on the other hand, we compute separately each term appearing
in Eq.~(\ref{Rq_ijkl}), for $(ijkl)=(1212)$:
\begin{enumerate}
\item
the term \ $-(q-1)(\bar f_{,11}\bar f_{,22}-\bar f_{,12}\bar f_{,21})$ \ equals
\begin{equation}
-(q-1)\left|\begin{array}{cc}
\bar f_{,11} & \bar f_{,12} \\
\bar f_{,21} & \bar f_{,22}
\end{array}\right|
\label{Rq1}
\end{equation}
which is proportional to $|H(\bar f)|$;
\item
the term $- \frac 14 \,g^{(q)\,st}\, (m_{s,22}m_{t,11}-m_{s,21}m_{t,12})$ equals
$1/(4g^{(q)})$ times
\begin{equation}
\left|\begin{array}{ccc}
m_{1,1}& m_{1,2}& m_{2,2} \\ m_{1,11}& m_{1,12}& m_{1,22}\\ m_{1,12}& m_{1,22}& m_{2,22}
\end{array}\right| + (q-1)
\left|\begin{array}{ccc} \bar f_{,1}^2&\bar f_{,1}\bar f_{,2}&\bar f_{,2}^2 \\
m_{1,11}& m_{1,12}& m_{1,22}\\ m_{1,12}& m_{1,22}& m_{2,22} \end{array}\right|
\label{Rq2}
\end{equation}
\item
the term $-(q-1)\frac 12 \,g^{(q)\,st} \left[ \bar f_{,s} (m_{t,11}\bar
f_{,22}-m_{t,12}\bar f_{,21}) +\right.$ $\left.+ \bar f_{,t}(m_{s,22}\bar
f_{,11}-m_{s,21}\bar f_{,12})\right]$ equals $(q-1)/(2g^{(q)})$ times
\begin{eqnarray}
\left|\begin{array}{cc}
\bar f_{,1} & \bar f_{,2} \\ m_{2,1} & m_{2,2}
\end{array}\right| \left( \left|\begin{array}{cc}
\bar f_{,11} & \bar f_{,12} \\ m_{1,21} & m_{1,22}
\end{array}\right|-\left|\begin{array}{cc}
\bar f_{,21} & \bar f_{,22} \\ m_{1,11} & m_{1,12}
\end{array}\right| \right) - && \nonumber \\
\! -\left|\begin{array}{cc}
\bar f_{,1} & \bar f_{,2} \\ m_{1,1} & m_{1,2}
\end{array}\right| \left( \left|\begin{array}{cc}
\bar f_{,11} & \bar f_{,12} \\ m_{2,21} & m_{2,22}
\end{array}\right|-\left|\begin{array}{cc}
\bar f_{,21} & \bar f_{,22} \\ m_{2,11} & m_{2,12}
\end{array}\right| \right) &&
\label{Rq3}
\end{eqnarray}
\item
and the term \mbox{$-(q-1)^2(\bar f_{,11}\bar f_{,22}-\bar f_{,12}\bar
f_{,21})$} $\, g^{(q)\,st}\,\bar f_{,s}\bar f_{,t}$ equals $(q-1)^2/g^{(q)}$
times
\begin{equation}
\left|\begin{array}{cc}
\bar f_{,11} & \bar f_{,12} \\
\bar f_{,21} & \bar f_{,22}
\end{array}\right| \,
\left\{\bar f_{,1} \left|\begin{array}{cc}
\bar f_{,1} & \bar f_{,2} \\ m_{2,1} & m_{2,2}
\end{array}\right|-\bar f_{,2} \left|\begin{array}{cc}
\bar f_{,1} & \bar f_{,2} \\ m_{1,1} & m_{1,2}
\end{array}\right|\,\right\}
\label{Rq4}
\end{equation}
\end{enumerate}
Thus $R^{(q)}_{1212}$ is obtained as the sum of the contributions
in expressions~(\ref{Rq1})--(\ref{Rq4}), %Rq1,Rq2,Rq3,Rq4
with their corresponding prefactors, and with $g^{(q)}$ given by
Eq~(\ref{detgq}). Notice that we have arranged all the second order
determinants as Jacobian ones, involving the two-parameters functions $m_1$,
$m_2$, and $\bar f$ as well as their first derivatives with respect to both
parameters. For brevity, we will denote the Jacobian determinant
$\partial(\phi,\psi)/\partial(\beta^1,\beta^2)$ simply as $J(\phi,\psi)$.
Also, the Hessian determinant of $\bar f$ is to be denoted as $H(\bar f)$
wherever it appears (see Eq.~(\ref{Rq1})~). Summing up, the {\it generalized
scalar curvature} of 2D-space becomes
\begin{widetext}
\begin{eqnarray}
R^{(q)}&=&\frac{2}{(g^{(q)})^2}\bigg\{\frac 14
\left|\begin{array}{ccc}
m_{1,1}& m_{1,2}& m_{2,2} \\ m_{1,11}& m_{1,12}& m_{1,22}\\ m_{1,12}& m_{1,22}& m_{2,22}
\end{array}\right|
+(q-1)\bigg[\frac 14
\left|\begin{array}{ccc}
\bar f_{,1}^2&\bar f_{,1}\bar f_{,2}&\bar f_{,2}^2 \\m_{1,11}& m_{1,12}& m_{1,22}\\ m_{1,12}& m_{1,22}& m_{2,22}
\end{array}\right|-J(m_1,m_2)\,|H(\bar f)|+ \nonumber\\
&& +\frac 12\left[J(\bar f,m_2)\left(J(\bar f_{,1},m_{1,2})-J(\bar f_{,2},m_{1,1})\right)-
 J(\bar f,m_1)\left(J(\bar f_{,1},m_{2,2})-J(\bar f_{,2},m_{2,1})\right)\right]\bigg]\bigg\}
\label{Rq}
\end{eqnarray}
\end{widetext}
with
\begin{equation}
g^{(q)}=J(m_1,m_2)+(q-1)\left[\bar f_{,1}\,J(\bar f,m_2)-\bar f_{,2}\,J(\bar
f,m_1)\right]
\end{equation}
In these expressions it is not difficult to recognize the extensive limit,
that coincides with the result given, for instance, in
Ref.~\onlinecite{jm_pra39}. Finally, one could regroup all the contributions
to $R^{(q)}$ and analyze the ensuing departure from the extensive situation,
which may eventually shed some light on the behavior of the system under
study. Those terms that do not contain any explicit factor $(q-1)^n$, or
``zeroth-order terms", are clearly the ones that survive in the limit
$q\rightarrow 1$, while higher order terms in $(q-1)$ represent the
successive corrections to the extensive case, when $q$ is close to one.

\section{The ideal gas case}
\label{section_idealgas}

We address here a classical ideal system consisting of $N$ free particles in
contact with a heat bath at temperature $T$ and pressure $P$ through a
movable wall (the volume is variable). Considering the $P$-$T$, or
Boguslavski, distribution, the standard partition function (usually denoted
as $Y_N(P,T)$ in textbooks) is given by
\begin{eqnarray}
Z_1(\alpha,\beta)&=&
\frac{1}{N!h^{3N}}\int_0^{\infty} dV
\int d^{3N}\!q\,d^{3N}\!p \ \mathrm{e}^{-\alpha V-\beta H} \nonumber\\
&=&
\alpha^{-(N+1)} \left(\frac{2\pi m}{h^2\beta}\right)^{3N/2}
\label{Z1idealgas}
\end{eqnarray}
where $H$ denotes the free particle Hamiltonian for $N$ particles
of mass $m$, and the intensive parameters are taken to be
$\alpha=P/kT$ and $\beta=1/kT$.

One can employ Eq.~(\ref{Z1idealgas}) as a starting point to
calculate the nonextensive thermodynamical magnitudes of the
system, by recourse to integral representations based on the
definition of the Euler gamma function (see, for instance, the
appendix in Ref.~\onlinecite{mppp_pa332} for details and a
discussion regarding its validity). Performing the so-called
Hilhorst transform, which is applicable if the nonextensivity
index $q$ is greater than 1, the generalized pseudo-partition
function can be written in the form
\begin{eqnarray}
\bar
Z_q(\alpha,\beta) & = & Z_1(\alpha,\beta) \times \nonumber \\
& \times &\frac{
\Gamma\left(\frac{1}{q-1}-\frac 52 N-1\right) \
\left(\frac{1}{q-1}\right)^{\frac 52 N+1}
}
{
\Gamma\left(\frac{1}{q-1}\right) \
[1-(q-1)I_q]^{\frac{1}{q-1}-\frac 52 N-1}
}
\label{Zqrayaidealgas}
\end{eqnarray}
where for brevity we have defined the quantity $I_q=\alpha\langle
V\rangle_q+\beta\langle H\rangle_q\equiv\alpha V_q+\beta U_q$, which is a
convenient combination of the intensive parameters and corresponding
generalized mean values for volume and energy, respectively. The range of
$q$ for which Eq.~(\ref{Zqrayaidealgas}) is valid is given by $0<q-1<(\frac
52 N+1)^{-1}$. Performing the same sort of real integral representation as
before, one can compute the $q$-expectation values of the relevant
operators. The following interesting result ensues
\begin{equation}
V_q= \frac{N+1}{\alpha} \qquad \mathrm{and} \qquad U_q= \frac{3N}{2\beta}
\label{vquq}
\end{equation}
while $I_q=\frac 52 N+1$. It is remarkable that these results are
independent of the value of the index $q$ (as long as it belongs to the
range of validity of the integral representation employed). Also, for later
purposes, we notice that $U_q$ does not depend on the pressure, while $V_q$
is a function of $\alpha$ only. In both expressions, $N$ is the (fixed)
number of particles that constitute the system under study; note that in the
``equation of state" there is a factor $N+1$ instead of $N$ as we have not
taken the thermodynamic limit which allows for that replacement.

\hfill

With these ingredients we can compute the generalized metric tensor and
other geometrical quantities of interest defined in
Section~\ref{section_infogeomq}. From Eqs.~(\ref{Zqrayaidealgas}) and
(\ref{vquq}), for the $q$-metric tensor (\ref{gqijZ}) we obtain the
following $2\times 2$ symmetric matrix
\begin{equation}\label{gqigas}
\left(g^{(q)}_{ij}\right) =
\bigg( \begin{array}{cc}
\frac{N+1}{\alpha^2} & 0 \\ 0 & \frac{3N}{2\beta^2}
\end{array}\bigg)-(q-1)\bigg(\begin{array}{cc}
\left(\frac{N+1}{\alpha}\right)^2 &
\quad \frac{3N(N+1)}{2\alpha\beta} \\
\frac{3N(N+1)}{2\alpha\beta} &
\quad \left(\frac{3N}{2\beta}\right)^2
\end{array}\bigg)
\end{equation}
for $1<q<1+(\frac 52 N+1)^{-1}$. The limit $q\rightarrow 1^+$ gives the
correct diagonal result already obtained by other authors
(Refs.~\onlinecite{j_jpa23} and~\onlinecite{br_pre51}, for instance). Here
the changes introduced by the nonextensive treatment appear as contributions
of order $(q-1)$: a dependence with $N^2$ appears since we have not
considered thermodynamic magnitudes per particle (as Brody and Rivier, for
instance, do). The $2 \times 2$-determinant of the matrix~(\ref{gqigas})
acquires a simple form in terms of the intensive parameters $\alpha$ and
$\beta$, and the number of particles $N$:
\begin{equation}
g^{(q)} = \frac{3N(N+1)}{2\alpha^2\beta^2 }\,[1-(q-1)(\frac52 N+1)]
\end{equation}
We compute explicitly the expressions in (\ref{Rq1})--(\ref{Rq4}) %Rq1,Rq2,Rq3,Rq4
(with their corresponding prefactors) and arrive, respectively, at
\begin{eqnarray}
\frac{(q-1)}{g^{(q)}} \left(\frac{3N(N+1)}{2\alpha^2\beta^2}\right)^2
& \times &
\{ -1+(q-1)(\frac 52 N+1) \}
\nonumber\\
\frac{(q-1)}{g^{(q)}} \left(\frac{3N(N+1)}{2\alpha^2\beta^2}\right)^2
& \times &
\{ -1 \} \nonumber
\\
\frac{(q-1)}{g^{(q)}} \left(\frac{3N(N+1)}{2\alpha^2\beta^2}\right)^2
& \times &
\{ 2 \} \nonumber
\\
\frac{(q-1)}{g^{(q)}} \left(\frac{3N(N+1)}{2\alpha^2\beta^2}\right)^2
& \times &
\{ -(q-1)(\frac 52 N+1) \}
\end{eqnarray}
Adding these four terms and multiplying by the factor $2/g^{(q)}$ we are led
to $R^{(q)}$. The computation of the generalized scalar curvature in this
example yields the outcome: $R^{(q)}=0$. It is noticeable that this result
(i) is independent of the intensive parameters $\alpha$ and $\beta$, and
(ii) is also independent of the nonextensivity index $q$. We can then assert
that in this case the outcome for the curvature in our geometrical approach,
performed within the framework of a nonextensive statistical setting, agrees
with the developments obtained by different authors within the structures of
the extensive formalism --see, for
instance,~Refs.~\cite{j_jpa23,br_pre51}--.

\section{Concluding remarks}
\label{section_conclusions}

We have discussed geometrical aspects of the space of parameters for a
physical system, within the framework of a nonextensive statistical setting
of fixed index $q$. For that purpose, we have employed a generalization of
the quantum Kullback--Leibler divergence as information gain. In particular,
we have derived the pertinent generalized metric tensor which is connected
with Fisher's information matrix. Our main findings in the present effort
are represented by Eqs.~(\ref{gqijZ}) and~(\ref{Rq}), which pave the way for
deriving the metric and curvature tensors from a generalized partition
function for a fixed value of the nonextensivity index.

The studies addressed here are relevant in the investigation of
critical behavior and phase transitions in thermodynamic systems.
Work on such sort of applications for interacting systems, using
generalized information measures, is currently in progress.

Finally, we remark that the use of alternative measures of distance between
pairs of density operators in the space of thermodynamic parameters could
yield an alternative method --perhaps a more sensible one-- to detect
thermodynamic changes in certain physical systems applying the geometric
machinery discussed in the present effort. To mention just one such
possibility, the Jensen--Shannon (JS) divergence can be used to give a
measure of distance between two probability distributions. A generalization
of the JS divergence in the framework of Tsallis statistical mechanics has
been investigated by Lamberti and co-workers \cite{lm_pa329,mlp_pa344}, who
obtained a monoparametric family of metrics.
This and other alternatives will be the subject of future work.

\begin{acknowledgments}

The authors acknowledge financial support from CONICET and UNLP,
Argentina. MP also acknowledges ANPCyT (PICT No.~03-11903/2002),
Argentina.

\end{acknowledgments}

\end{document}